\documentclass[10pt,preprintnumbers,pra,aps,
nofootinbib,tightenlines,floatfix,superscriptaddress,
longbibliography,notitlepage,twocolumn]{revtex4-2}

\usepackage[colorlinks=true]{hyperref}
\hypersetup{
  linkcolor  = TealBlue!90!Black,
  citecolor  = YellowGreen!90!Black,
  urlcolor   = Peach!90!Black,
  colorlinks = true,
}
\usepackage{amsmath,amssymb}
\usepackage{graphicx}
\usepackage[utf8]{inputenc}
\usepackage{csquotes}
\usepackage{bm}
\usepackage[dvipsnames]{xcolor}
\usepackage[margin=0.5in, footskip = 0.5in]{geometry}
\usepackage{setspace}
\allowdisplaybreaks
\newcommand\scalemath[2]{\scalebox{#1}{\mbox{\ensuremath{\displaystyle #2}}}}

\newcount\hour \newcount\hourminute \newcount\minute
\hour=\time \divide \hour by 60
\hourminute=\hour \multiply \hourminute by 60
\minute=\time \advance \minute by -\hourminute

\newcommand\CG[6]{\ensuremath{\langle #1, #2, #3,#4 | #5, #6\rangle }}
\newcommand\CGstar[6]{\ensuremath{\langle #5, #6 | #1, #2, #3, #4\rangle }}

\begin{document}

\title{\texorpdfstring{Some Conceptual Aspects of Operator Design \\ for Quantum Simulations of Non-Abelian Lattice Gauge Theories \\
\vspace{0.1cm}
\begin{minipage}{0.7\textwidth}
\centering
\textmd{\small{\it Contribution to proceedings of the 2021 Quantum Simulation for Strong Interactions (QuaSi) Workshops~\cite{quasiproceedingsSite} at the InQubator for Quantum Simulation (IQuS)}}
\end{minipage}
\vspace{-0.1cm}
}{Some Conceptual Aspects of Operator Design for Quantum Simulations of Non-Abelian Lattice Gauge Theories}
}
\author{Anthony Ciavarella}
\email{aciavare@uw.edu}
\affiliation{InQubator for Quantum Simulation (IQuS), Department of Physics, University of Washington, Seattle, WA 98195, USA}
\author{Natalie Klco}
\email{natklco@caltech.edu}
\thanks{speaker}
\affiliation{Institute for Quantum Information and Matter (IQIM) and Walter Burke Institute for Theoretical Physics, California Institute of Technology, Pasadena CA 91125, USA}
\author{Martin J.~Savage}
\email{mjs5@uw.edu}
\affiliation{InQubator for Quantum Simulation (IQuS), Department of Physics, University of Washington, Seattle, WA 98195, USA}

\begin{abstract}
  In the Kogut-Susskind formulation of lattice gauge theories, a set of quantum numbers resides at the ends of each link to characterize the vertex-local gauge field.
  We discuss the role of these quantum numbers in propagating correlations and supporting entanglement that ensures each vertex remains gauge invariant, despite time evolution induced by operators with (only) partial access to each vertex  Hilbert space.
  Applied to recent proposals for eliminating vertex-local Hilbert spaces in quantum simulation, we describe how the required entanglement is generated via delocalization of the time evolution operator with nearest-neighbor controls.
 These hybridizations, organized with qudits or qubits, exchange classical operator preprocessing for reductions in quantum resource requirements that extend throughout the lattice volume.
\end{abstract}
\preprint{IQuS@UW-21-024}
\maketitle

\section{Introduction}

Remarkable progress has been made in conceptual and quantitative understanding of Standard Model physics 
using classical computational architectures (e.g., through lattice QCD calculations~\cite{Bazavov:2019lgz,Aoki:2019cca,Joo:2019byq}).
However, dynamical properties of entangled quantum many-body systems, at scale with a complete quantification of uncertainties, generally lie beyond the capabilities of classical computing.
Pursuing important scientific objectives at the frontiers of complexity and entanglement inspires a restoration of natural processes into computational frameworks, i.e., incorporating quantum systems directly~\cite{5392446,Manin1980,Benioff:1980,Fredkin1982,Feynman1982,Lloyd1073}.
With motivations from fundamental science to large-scale quantum computation, and following developments pioneered in analog quantum simulators~\cite{Zohar:2012ay,Tagliacozzo:2012df,Banerjee:2012pg,Zohar:2012xf,Banerjee:2012xg}, 
progress toward this goal continues from diverse research programmes focusing upon the implementation and codesign of lattice gauge theories (LGTs) with present and future quantum devices.  

It is anticipated that, 
when parallelization of quantum gate implementations is combined with descriptions of quantum dynamics in terms of local operators, 
quantum simulations can be performed with computational times (wall clock) that scale polynomially with the duration of simulated-time evolution~\cite{Feynman1982,Lloyd1073}.
Beyond resonating with physical expectations, a local operator framework provides a clear technique for organizing quantum simulation design.
In order for LGTs with local symmetries to be captured by the dynamics of local operators, 
the dimensionality of the  Hilbert space used for  simulation is exponentially larger than the gauge-invariant space within which a simulation should remain.
Throughout the array of gauge field simulation designs discussed in this {\it QuaSi} workshop, 
significant focus has been placed upon strategies  to minimize and reliably avoid such excess Hilbert space(s).

In the following discussion, 
we consider quantum simulations of LGTs using a
local multiplet basis in which vertex-local quantum degrees of freedom 
are {\it integrated out} to reduce the footprint of the associated gauge-variant Hilbert space~\cite{Banuls:2017ena,Klco:2019evd,Ciavarella:2021nmj}.
Extending the presentation in Ref.~\cite{Ciavarella:2021nmj} of scalability and implementation of truncated SU(3) dynamics in small volumes 
compatible with the local operator's extent on IBM's superconducting quantum devices, 
we further discuss the proposed hybrid design of local operators.
In particular, we explicitly illustrate how local color-isospin and 
hypercharge degrees of freedom provide an underlying entangled fabric protecting Gauss's law throughout the lattice, 
and how this fabric can be removed when local operators are classically processed to gain full access to local vertex Hilbert spaces.

\section{Multiplet Basis Hilbert Space}
The Hamiltonian formulation of LGTs is described by~\cite{KogutSusskind1975}
\begin{multline}
\hat H  =
\frac{g^2}{2 a^{d-2}}
\sum_{ b, {\rm links}}
 |  \hat {\bf E}^{(b)} |^2 \\
 -
\frac{1}{2 a^{4-d} g^2} \sum_{{\rm plaquettes} }
\left[ \hat { \Box}  + \hat {\Box}^\dagger
\right]
\ \ \ ,
\label{eq:QCDham}
\end{multline}
where $a$ is the lattice spacing, $g$ is the strong coupling constant, $d$ is the number of spatial dimensions, and conventional constant factors have been omitted.
In the electric representation basis of SU(3) LGT, the lattice Hilbert space may be regarded as a tensor product of link Hilbert spaces each carrying eight quantum numbers of the form $|p, q\rangle |T_\ell, T_\ell^z, Y_\ell\rangle |T_r, T_r^z, Y_r\rangle$, 
where the first two define the irreducible representation 
and the following six represent the color-isospin and color-hypercharge localized to the left and right side of the link.
It is in this basis, with eight quantum registers dedicated to each link, that the theoretical efficiency of Yang-Mills LGT quantum simulation was first established~\cite{Byrnes:2005qx}.
The electric terms consist of diagonal operators at each link
with matrix elements characterized by quadratic Casimirs of the associated irreducible representation.
The magnetic terms consist of operators that act upon four-link plaquettes of the lattice
\begin{multline}
\hat {\Box}   =
{\rm Tr}\left[\
\hat U^{\bf 3}({\bf x}, {\bf x} + a {\bm \mu}) \right.
\hat U^{\bf 3}({\bf x} + a {\bm \mu}, {\bf x} + a {\bm \mu}+ a {\bm \nu})
\\
\hat U^{\bf 3}({\bf x} + a {\bm \mu}+ a {\bm \nu}, {\bf x} + a {\bm \nu})
\left.
\hat U^{\bf 3}( {\bf x} + a {\bm \nu}, {\bf x})\
\right]
\ \ \ ,
\label{eq:PlaqOp}
\end{multline}
where  $\hat U^{\bf 3}({\bf x},{\bf y})$ are link operators in the fundamental representation of SU(3), and ${\bm x}$, ${\bm\mu}$ and ${\bm\nu}$ are position and unit vectors (implicit on left) that define the location and orientation of the plaquette.

The plaquette operators provide the maximally localized structure capable of generating excitations of the field while maintaining gauge invariance throughout the lattice.
In the following, we discuss the impact on quantum simulation design of integrating the local quantum numbers, $|T, T^z, Y\rangle^{\otimes 2d}$, at each spatial vertex~\cite{Banuls:2017ena,Klco:2019evd,Ciavarella:2021nmj}~\footnote{Throughout this work, the collection of local quantum numbers, $|T, T^z, Y\rangle$, will be notated by a single plain-type variable that indexes the physical states in the associated irreducible representation of dimension $  \dim(p,q) = \frac{1}{2} (p+1)(q+1)(p+q+2)$.}.
Though no portion of the Hilbert space is dedicated to their representation after local integration, the impact of these local quantum numbers is incorporated in the design of magnetic time evolution operators.
As a result, this hybrid approach with classical preprocessing dramatically reduces the quantum degrees of freedom necessary to simulate the field (and thus the ratio of gauge variant to gauge invariant Hilbert space) at the cost of minimally delocalizing the plaquette operator with controls on the nearest neighbor links.

\section{Vertex-Local Quantum Numbers}
When all eight quantum numbers are retained in each link Hilbert space, the preservation of Gauss's law upon application of the plaquette operator can be interpreted to arise through destructive interference of Gauss-law-violating configurations.
For a simple example, consider a pair of vertices that have been excited in the fundamental representation, 
$\mathbf{3}$, passing through a vertex where all other links reside in the singlet state.
With the link operator generating local superpositions of viable excitations on each link as~\cite{KogutSusskind1975,Zohar:2014qma},
\begin{multline}
  \hat U_{\alpha, \beta}^{\mathbf{3}} |\mathbf{R},a, b\rangle
  = \sum_{\oplus {\bf R}', \vec{\Gamma}} \sum_{a' b'} \sqrt{\frac{\dim( \mathbf{R})}{\dim(\mathbf{R}')}}  |\mathbf{R}', a', b'\rangle \\
  \CG{\mathbf{R}}{a}{\mathbf{3}}{\alpha}{\mathbf{R}'}{a'}_{\Gamma_1}
  \CGstar{\mathbf{R}}{b}{\mathbf{3}}{\beta}{\mathbf{R}'}{b'}_{\Gamma_2} \ \ \ ,
  \label{eq:linkop}
\end{multline}
applying a further plaquette operator incorporating the excited pair produces content in
$\mathbf{3} \otimes \mathbf{3} = \bar{\mathbf{3}} \oplus \mathbf{6}$.
The vertex wavefunction, 
$|\psi\rangle_v = \frac{1}{3} \sum_s |\mathbf{3}, \ell_1, s\rangle |\mathbf{3}, s, r_2\rangle |\mathbf{1}, 0, 0\rangle^{\otimes 2d-2}$, transitions to
\begin{multline}
  \frac{1}{3}\sum_{\beta, s} 
  \hat U_{\alpha \beta}^{\mathbf{3}} 
  |\mathbf{3}, \ell_1, s\rangle 
  \hat U_{\beta \gamma}^{\mathbf{3}} 
  |\mathbf{3}, s, r_2\rangle |\mathbf{1}, 0, 0\rangle^{\otimes 2d-2} = \frac{1}{3}\sum_{\beta, s} \\
  \scalemath{0.82}{
  \left( \sum_{\mathbf{R}'_1,\ell_1', r_1'} \sqrt{\frac{\dim(\mathbf{3})}{\dim(\mathbf{R}'_1)}} |\mathbf{R}'_1, \ell_1', r_1'\rangle
  \CG{\mathbf{3}}{\ell_1}{\mathbf{3}}{\alpha}{\mathbf{R}'_1}{\ell_1'}
  \CGstar{\mathbf{3}}{s}{\mathbf{3}}{\beta}{\mathbf{R}'_1}{r_1'}
  \right)}\\
  \scalemath{0.82}{
  \left(\sum_{\mathbf{R}'_2, \ell_2', r_2'} \sqrt{\frac{\dim( \mathbf{3})}{\dim(\mathbf{R}'_2)}} |\mathbf{R}'_2, \ell_2', r_2'\rangle
  \CG{\mathbf{3}}{s}{\mathbf{3}}{\beta}{\mathbf{R}'_2}{\ell_2'}
  \CGstar{\mathbf{3}}{r_2}{\mathbf{3}}{\gamma}{\mathbf{R}'_2}{r_2'}
  \right)}
  \\* \otimes |\mathbf{1}, 0, 0\rangle^{\otimes 2d-2} \ \ \ ,
  \label{eq:vertex1}
\end{multline}
for $\mathbf{R}'_{1,2} \in \{\bar{\mathbf{3}}, \mathbf{6}\}$.
If these two superpositions were independent, Gauss's law would be violated as non-zero amplitude would be produced for a $\bar{\mathbf{3}} \rightarrow \mathbf{6}$ transition at the vertex, and thus the generation of color flux.
However, the contraction of link operators across vertices, in this case the interference described by the Clebsch-Gordan (CG) sum
\begin{equation}
  \sum_{\beta, s} \CGstar{\mathbf{3}}{s}{\mathbf{3}}{\beta}{\mathbf{R}'_1}{r_1'} \CG{\mathbf{3}}{s}{\mathbf{3}}{\beta}{\mathbf{R}'_2}{\ell_2'} = \delta_{\mathbf{R}'_1, \mathbf{R}'_2 } \delta_{r_1', \ell_2'} \ \ \ ,
\end{equation}
provides the necessary correlation to remove population from unphysical regimes of the Hilbert space.
The final vertex wavefunction contains only population with zero color flux entering/exiting the active area of the plaquette, such that Eq.~\eqref{eq:vertex1} becomes
\begin{multline}
  \hat U_{\alpha \beta}^{\mathbf{3}} 
  \hat U_{\beta \gamma}^{\mathbf{3}} 
  |\psi\rangle_v =  \sum_{\mathbf{R}', s', \ell_1', r_2'} \frac{1}{\dim(\mathbf{R}')} \\ |\mathbf{R}', \ell_1', s'\rangle |\mathbf{R}', s', r_2'\rangle |\mathbf{1}, 0, 0\rangle^{\otimes 2d-2} \\
  \CG{\mathbf{3}}{\ell_1}{\mathbf{3}}{\alpha}{\mathbf{R}'}{\ell_1'}
  \CGstar{\mathbf{3}}{r_2}{\mathbf{3}}{\gamma}{\mathbf{R}'}{r_2'} 
  \ \ \ ,
\end{multline}
with normalization indicating the higher probability of generating the lower-energy flux reversal of the $\bar{\mathbf{3}}$ configuration than the production of additional flux for the $\mathbf{6}$ configuration.

When the second plaquette operator is instead placed at the vertex along a different axis sharing one excited link in $|\psi\rangle_v$, correlations among the projection quantum numbers begin to produce the vertex CG factors,
\begin{multline}
  \frac{1}{3} \sum_{\alpha,r_1} 
  |\mathbf{3}, \ell_1, r_1\rangle 
  \hat U_{\alpha, \beta}^{\mathbf{3}} 
  |\mathbf{3}, r_1, r_2\rangle 
  \hat U^{\mathbf{3\ }\dagger}_{\delta, \alpha} |\mathbf{1}, 0, 0\rangle  \sim \\ \frac{1}{3} \sum_{\mathbf{R}_2',r_1, \ell_2', \ell_3'}\frac{1}{\sqrt{\dim (\mathbf{R}_2')}} \langle \mathbf{3}, r_1, \mathbf{3}, \ell_3'|\mathbf{R}_2', \ell_2'\rangle  \\ |\mathbf{3}, \ell_1, r_1\rangle  |\mathbf{R}_2', \ell_2', r_2'\rangle  |\bar{\mathbf{3}}, \ell_3', r_3'\rangle \ \ \ ,
\end{multline}
where the $r_{2,3}$ factors are notationally neglected as features of the lattice wavefunction external to the vertex.
Once again, the  configuration with lower electric energy ($\epsilon^{abc}$-contracted) is seen to have an enhanced amplitude over the generation of additional color flux on the second link.
In some sense, the qutrit-type maximally-entangled state that spans two links of the vertex after application of the first plaquette becomes distributed as a three-link singlet contraction upon implementation of a neighboring plaquette operator.
This process describes how such vertex CGs are sequentially constructed for the preservation of Gauss's law despite the plaquette operator only acting on two-link subspaces of a vertex.
Continuation of this process explores only the gauge-invariant subspace and generates CG vertex factors 
that do not contain $\mathbf{3}$ or $\bar {\mathbf{3}}$ irreps (or those that define the plaquette operator) upon employment of higher-order CG identities.
For example,
\begin{multline}
  \sum_{r_1, \ell_2, \ell_3, \beta} 
  \hat U_{\alpha, \beta}^{\mathbf{3}} |\mathbf{3}, \ell_1, r_1\rangle 
  \hat U_{\beta, \gamma}^{\mathbf{3}} | \mathbf{3}, \ell_2, r_2\rangle\\ \otimes | \mathbf{8}, \ell_3, r_3\rangle \langle \mathbf{3}, r_1 | \mathbf{8}, \ell_3, \mathbf{3}, \ell_2\rangle  \sim \\ \sum_{r_1, r_1', \ell_2, \ell_2', \beta}   \sqrt{\frac{\dim(\mathbf{3})}{\dim(\mathbf{R}_1')\dim(\mathbf{R}_2')} }
  \CGstar{\mathbf{3}}{r_1}{\mathbf{3}}{\beta}{\mathbf{R}_1'}{r_1'}
  \\
  \CG{\mathbf{3}}{\ell_2}{\mathbf{3}}{\beta}{\mathbf{R}_2'}{\ell_2'}
  \CGstar{\mathbf{8}}{\ell_3}{\mathbf{3}}{\ell_2}{\mathbf{3}}{r_1}
  \\
  |\mathbf{R}_1', \ell_1', r_1'\rangle |\mathbf{R}_2', \ell_2', r_2'\rangle |\mathbf{8}, \ell_3, r_3\rangle       \ \ \ ,
\end{multline}
can be seen to produce the $\mathbf{6}$-$\mathbf{6}$-$\mathbf{8}$ vertex factor,
\begin{multline}
 \mathcal{N}_{\mathbf{R}_1', \mathbf{R}_2'}
  \CGstar{\mathbf{8}}{\ell_3}{\mathbf{R}_2'}{\ell_2'}{\mathbf{R}_1'}{r_1'} =
  \sum_{\beta, r_1, \ell_2}
  \CGstar{\mathbf{3}}{r_1}{\mathbf{3}}{\beta}{\mathbf{R}_1'}{r_1'}
  \\
  \CG{\mathbf{3}}{\ell_2}{\mathbf{3}}{\beta}{\mathbf{R}_2'}{\ell_2'}
  \CGstar{\mathbf{8}}{\ell_3}{\mathbf{3}}{\ell_2}{\mathbf{3}}{r_1} \ \ \ ,
\end{multline}
with coefficients (in a particular phase convention) of $\mathcal{N}_{\bar{\mathbf{3}},\bar{\mathbf{3}}} = \frac{1}{2}$, 
$ \mathcal{N}_{\mathbf{6}, \mathbf{6}} = \frac{\sqrt{10}}{4}$, 
$\mathcal{N}_{\bar{\mathbf{3}}, \mathbf{6}} = \frac{\sqrt{3}}{2}$, 
and $\mathcal{N}_{\mathbf{6}, \bar{\mathbf{3}}}= \frac{\sqrt{3}}{2\sqrt{2}} $.
Note that the asymmetry between $\mathcal{N}_{\bar{\mathbf{3}}, \mathbf{6}}$ and $\mathcal{N}_{\mathbf{6}, \bar{\mathbf{3}}}$ is 
not physical but arises to compensate for an asymmetry in CG notation e.g., normalizing $\langle \mathbf{1}, \mathbf{3} |\mathbf{3} \rangle$ and $\langle \mathbf{1}|\mathbf{3}, \bar{\mathbf{3}}\rangle$ with a relative factor of $\sqrt{3}$, and that indexing of conjugate representations has been chosen such that $\langle \mathbf{3}, i, \bar{\mathbf{3}}, j|1\rangle\propto \delta_{i, j}$.
The above demonstration emphasizes the role of local projection quantum numbers in tracking and distributing correlations throughout the lattice.
These embedded correlations allow localized operators---accessing only a subset of the vertex Hilbert space---to preserve the gauge symmetry.
In particular, the high-dimensional CG factors present in physical states far from the strong coupling vacuum may be interpreted as elaborately woven contractions from the interaction of neighboring plaquettes

It is worth noting that there is potential for leveraging this mechanism for the efficient quantum calculation of high-order CG factors as perturbative quantum simulation and final state projection on small lattices.
The following hybrid approach, developed in Refs.~\cite{Banuls:2017ena,Klco:2019evd,Ciavarella:2021nmj}, illustrates how the quantum simulation of Yang-Mills on large lattices may be designed in layers, allowing high-order CGs to be calculated independently (i.e., not requiring quantum coherence) and subsequently incorporated into a preprocessing of the local plaquette operator utilized for coherent time evolution.

\section{Local Integration}

While the local quantum numbers provide value in simplifying  the operators describing the theory, their cost for practical quantum simulation is significant.
As discussed in Ref.~\cite{Byrnes:2005qx}, if a truncation is placed on the number of tensor indices of the irreducible representation on each link, $p_{\rm max} = q_{\rm max} = \Lambda$, the isospin and hypercharge registers will have the following physical dimensions: $\left[ T \right] = 1 + 2 \Lambda$, $\left[ T^z\right]=1 + 4 \Lambda$, and $\left[ Y\right] = 1 + 6\Lambda$.
Assigning qubit registers to each of the eight quantum numbers of the link Hilbert space, leads to a qubit count of
\begin{equation}
  N_q^{\rm (link)} = 2 \left\lceil\log_2 \left(1+\Lambda\right) \right\rceil + 2 \sum_{j = 1}^3 \left\lceil \log_2\left( 1 + 2 j \Lambda\right) \right\rceil \ \ \ ,
\end{equation}
where the first(second) term accounts the qubits representing the irreducible representation(local quantum numbers).
While these two contributions to the link Hilbert space have similar asymptotic scalings, their prefactors produce dramatic practical consequences.
At all truncations, the number of qubits dedicated to the local quantum numbers exceeds that of the irrep, by a multiplicative factor scaling roughly as $\sim 3+\log_\Lambda(48)$.
At the lowest non-trivial truncation of $\Lambda = 1$, which includes local irreps of $\left\{ \mathbf{1}, \mathbf{3}, \bar{\mathbf{3}}, \mathbf{8}\right\}$, the two qubits dedicated to the irrep quantum numbers are overshadowed by the 16 qubits required to populate registers for the vertex-local quantum numbers.

The polynomial growth in the number of relevant CG factors, as well as the theoretical efficiency of their computation at fixed rank (quantum~\cite{2006PhRvL..97q0502B} or classical~\cite{Narayanan2006,de2006computation}), has inspired a hybrid modification to electric representation basis simulation techniques.
Extending the results of Refs.~\cite{Banuls:2017ena,Klco:2019evd} to the SU(3) gauge group, as discussed in Ref.~\cite{Ciavarella:2021nmj}, the responsibility of distributing correlations within each vertex to maintain gauge singlets can be transferred from the extra Hilbert space of local quantum numbers into the plaquette operator design.
In particular, the vertex CGs that would be generated via correlations among the $|T, T^z, Y\rangle$ Hilbert spaces may be incorporated into plaquette matrix elements themselves, leading to an operator acting in a reduced lattice Hilbert space of the form $|p, q\rangle$ on each link.

Because complete information of the local quantum number correlation structure is retained, embedded implicitly in the irrep configuration, this reduction does not affect the physical Hilbert space dimensionality.
However, that of the unphysical space is dramatically reduced.
Importantly, this redundancy removal does not eliminate the unphysical Hilbert space entirely.
In particular, a non-Abelian singlet constraint is still present at each vertex in the form
\begin{equation}
  \text{mod}_3(\mathbf{p}) = \text{mod}_3 (\mathbf{q})
\end{equation}
with
\begin{equation}
  \left(\mathbf{p}, \mathbf{q}\right) = \sum_{j \in \rm incoming} (p,q)_j + \sum_{j \in \rm outgoing} (q,p)_j  \ \ \ ,
\end{equation}
summed over all links at the vertex.
The remaining unphysical Hilbert space continues to serve its canonical role of allowing the theory to be described in terms of a small number of local operators.
The tradeoff of this local integration arises in the 1.) complexity of preprocessing and compiling the plaquette operator incorporating high-dimensional CG vertex factors and 2.) extension of the plaquette operator's spatial extent.
Fortunately for the former, these local operators need to be calculated and designed only once, after which they may be utilized extensively and throughout the volume for efficient implementation of time evolution.
Fortunately for the latter, the necessary extension remains at the scale of the lattice spacing and requires only controls extending to nearest neighbor links.
\begin{figure}
  \centering
  \includegraphics[width=0.9\columnwidth]{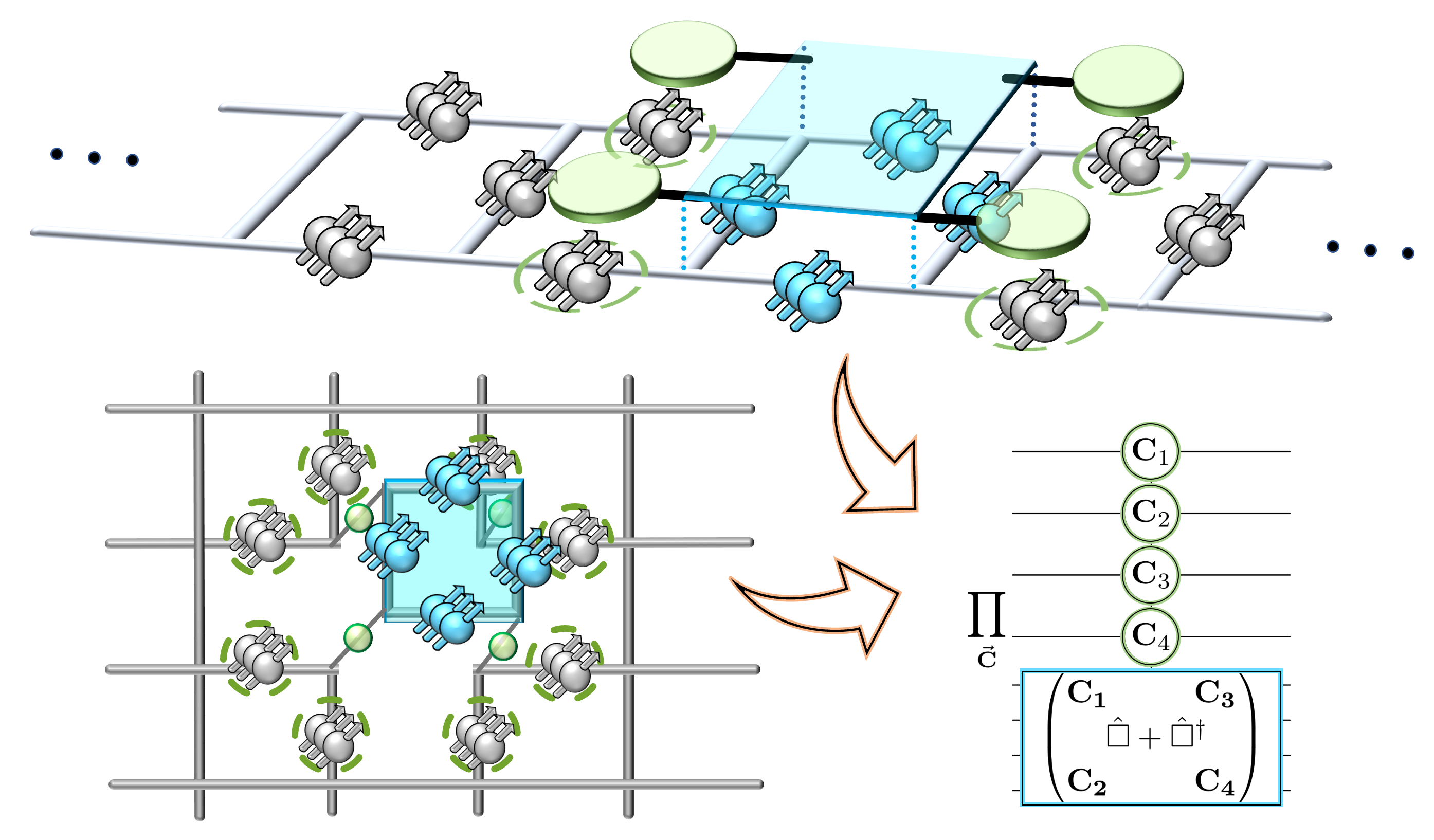}
  \caption{Structure of the plaquette operator upon integration of local quantum numbers in (top) a one-dimensional string of plaquettes~\cite{Klco:2019evd} and (bottom) two spatial dimensions.  The blue squares indicate the active quantum registers, the green circles indicate the neighboring controls, and the dashed green circles indicate the quantum registers upon which the controls depend.}
  \label{fig:plaquettediagram}
\end{figure}
As depicted in Fig~\ref{fig:plaquettediagram}, controls (green circles) are established on nearest neighbor links in all $D$ spatial directions.
The lower diagram of this figure emphasizes how the matrix elements calculated in the 1D plaquette string, as performed for SU(2) in Ref.~\cite{Klco:2019evd} and SU(3) in Ref.~\cite{Ciavarella:2021nmj}, are sufficient for defining the plaquette operator also in higher dimensions.
Whether this structure is utilized in practice through the introduction of an auxiliary \enquote{conduit} link as depicted or simply in effect via compilation, the controls depend only upon the total color flux entering/exiting the active Hilbert space at the vertex.

\section{Implementation}
Delocalizing operators in a quantum simulation protocol often leads to exponential difficulty in their compilation into a basic hardware gate set;
the extreme example of this being global bases~\cite{Ciavarella:2021nmj} in which gauge invariant Hilbert spaces are completely removed and hardware quantum states are mapped to physical configurations of the lattice volume.
With the proposed local integration strategy, the retained local operator structure and qudit framework allows clear organization of time evolution operators.
The electric operators are diagonal 1- or 2-qudit operators while the magnetic time evolution circuit may be decomposed according to the non-zero physical matrix elements of the plaquette operator.
As shown at the bottom right of Fig.~\ref{fig:plaquettediagram}, the plaquette operator can be first expanded in a product of control sectors, $\vec{C}$.
The operators in each control sector trivially commute, leading this product to be gauge invariant through Trotterized time evolution.
Operators in each control sector may be decomposed into Givens rotations such that each unitary operator is associated with a physical plaquette transition and the coefficients determined from gauge invariant matrix elements.
This compilation in terms of Givens rotations, while gauge invariant, introduces a source of systematic error upon Trotterization.
Ref.~\cite{Ciavarella:2021nmj} discusses in detail the  scaling of the number of physical matrix elements in the plaquette operator to quantify the Givens circuit depth required with this compilation approach.
While technically efficient---with a number of Givens rotations per plaquette operator scaling polynomially with the field truncation as $\mathcal{O}\left(\Lambda^{16}\right)$---the high degree of the polynomial scaling presents continued challenge, even for low energy wavefunctions that are expected to converge exponentially in field space.
However, experience with the impact of hardware and algorithmic co-design on anticipated quantum resources for quantum chemistry applications~\cite{Reiher2017} suggests ample opportunity for analogous refinements in the quantum simulation of field theories.

The gauge invariance of the designed qudit time evolution operator, which survives Trotterization, and the presence of residual unphysical Hilbert space provides a built-in mechanism for detecting local errors.
In particular, whether applying a non-destructive measurement of the Gauss's law operators~\cite{Stryker:2018efp} or projectively measuring the final quantum state (e.g., as performed in Ref.~\cite{Klco:2019evd}), post-selection into the gauge-invariant subspace provides reliable criteria for suppressing incoherent, vertex-density bit flip errors to $\mathcal{O}(p^2)$.
Though the presence of physical states at distance-2 upon local bit flips leads to incomplete availability of correction with current methods, the structure of the retained gauge symmetry allows passive detection of this category of error.
In light of the natural ability of gauge theories to protect distributed quantum degrees of freedom from local sources of quantum noise~\cite{Kitaev:1997wr,Gottesman:1997zz,Bravyi:1998sy},  further work incorporating natural error robustness
is at the frontier of gauge theory quantum simulation.

In terms of plaquette operator localization, this hybrid multiplet basis may be contextualized in the literature as intermediate between the structure of Ref.~\cite{Byrnes:2005qx} and that of the Schwinger bosons underlying prepotentials and the loop-string-hadron (LSH) formulation~\cite{Raychowdhury:2018osk,Raychowdhury:2019iki}.
While the former places all gauge field information at the link, the latter captures the field through gauge-invariant operators local to each vertex.
With the integrated multiplet basis, the projection quantum numbers are localized to each vertex  while the irrep quantum number remains at the link.
The subsequent local integration of the projection quantum numbers produces a nearest-neighbor delocalization of the same spatial extent as the plaquette operator in the LSH formulation.

\section{Closing Remarks}

With the quantum simulation of non-Abelian gauge theories relevant to Standard Model physics in its infancy, understanding and building upon an array of
small, low-dimensional LGTs is an essential part of present-day development, with implications beyond quantum field theories.
In the context of generating the entanglement necessary to satisfy local gauge constraints, we have discussed aspects of hybrid operator design  for a digital quantum simulation of the Kogut-Susskind formulation that trades-off a reduced Hilbert space for neighbor-controlled evolution operators and associated classical computation.

\begin{acknowledgments}
We thank the QuaSi community for engaging and valuable discussions, both at the QuaSi workshops and through the extensive literature that exceeds included references.
AC is supported in part by Fermi National Accelerator Laboratory PO No. 652197.
NK is supported in part by the Walter Burke Institute for Theoretical Physics, and by the U.S. Department of Energy Office of Science, Office of Advanced Scientific Computing Research, (DE-SC0020290), and Office of High Energy Physics DE-ACO2-07CH11359.
MJS is supported in part by the U.S. Department of Energy, Office of Science, Office of Nuclear Physics, Inqubator for Quantum Simulation (IQuS) under Award Number DOE (NP) Award DE-SC0020970.
\end{acknowledgments}

\bibliography{bibsyrae}

\end{document}